# Energy-flux pattern of inverse Goos—Hänchen shift in photonic crystals with negative index of refraction

Jinbing Hu[1], Binming Liang[1*], Jiabi Chen[1], QiangJiang[1], YanWang[2], Songlin Zhuang[1]

[1]Shanghai Key Lab of Modern Optical System, University of Shanghai for Science and Technology, No. 516 Jun Gong Road, Shanghai 200093, China

[2]the Institute of Physics and Communication Electronics, Jiangxi Normal University, Nanchang 330022, China.

*Corresponding authors: liangbinming@sina.com

The energy-flux patterns of inverse Goos—Hänchen (GH) shift around the interface between air and negatively refractive photonic crystal (NRPhC) with different surface terminations is investigated. Results show that NRPhC exhibits inverse GH shift in TM and TE polarization, and the localization and pattern of energy flux differ in TM and TE polarizations and are strongly affected by surface termination. This is different to the condition of negative-permittivity materials (i.e., metal), which only presents inverse GH shift in TM polarization. In the case of TE polarization, the energy—flux pattern exhibits the flux of backward wave whose localization changes from the surface to inside of NRPhC with the variation of surface termination. In the case of TM polarization, the energy—flux pattern is always confined within the surface of NRPhC, whereas its pattern changes from the flux of backward wave to vortices at the surface of NRPhC, which is different to the energy flux of TM polarization of metal. By properly truncating the surface of NRPhC we can control the magnitude of inverse GH shift for TM—and/or TE—polarized light.

Since experimental implementation of negative refraction in photonic crystals (PhC)[1, 2], PhC with negative index of refraction, i.e., negatively refractive photonic crystals (NRPhC), has been the subject of extensive interests because of its exceptional phenomena, such as inverse Doppler Effect[3, 4], inverse Goos-Hänchen effect[5, 6], *et al*. Especially, the experimental observation of inverse Doppler Effect in NRPhC at optical frequency[3] provides a basis for exploration of inverse GH shift in NRPhC. Inverse GH shift of NRPhC[5] is a phenomenon that occurs at the interface between air and NRPhC. And it was reported that the magnitude of inverse GH shift is strongly dependent on surface termination of NRPhC †:TM–polarized light is characterized by an inverse GH shift of more than twenties of lattices, whereas TE-polarized light exhibits infinitesimal inverse GH shift when the surface is terminated with the outermost row of air holes complete. However, an inverse condition occurs when the outermost row of air holes is cut in quarter or so. Furthermore, the maximum value of inverse GH shift of TE polarization is larger than that of TE polarization for same NRPhC structure.

The present paper aims to investigate the physical mechanism of the difference of inverse GH shift between TM and TE polarizations by studying the energy flux around the interface between air and NRPhC. To the best of our knowledge, no study has focused on this problem.

The first consideration of energy flux of inverse GH shift was reported by Lai[7]. When a TM-polarized Gaussian beam encounters the interface of two media with opposite permittivity there are closed-loop flux lines around the interface in addition to the incoming-outgoing flow pattern in the rest of the region. That is as if TM polarized incident beam seems to be reflected from the plane laying at a distance ahead the real interface. In his paper the second medium was assumed to be negative-permittivity material (i.e., metal), which is different to NRPhC due to the fact that NRPhC has remarkable surface effect, in addition to equivalent to negative-index materials at some frequencies. Besides, the GH shift and energy flux of TE polarization were not studied in his paper. There is, therefore, a need to study the energy flux of inverse GH shift around the NRPhC-air interface.

We study the same two–dimensional negatively refractive photonic crystals (2D–NRPhC) as the one considered By Notomi[8]. The 2D–NRPhC consists of hexagonal lattice of air holes in a dielectric background with a dielectric constant of 12.96 (e.g., GaAs or Si at 1,550 nm). The radius of air holes r is 0.4 *a*, where *a* is the lattice constant. We set the *xz* plane as the plane of periodicity, which is parallel to the propagation vector. The interface between air and 2D–NRPhC is parallel to the Γ—K direction. The effective index of 2D-NRPhC is nearly isotropic and has a negative value in normalized frequency regions of $0.29-0.34$ $\omega a/2\pi c$ and $0.51-0.56$ $\omega a/2\pi c$ for TE (i.e., the electric field is perpendicular to the plane of periodicity) and TM (i.e., the electric field is in the plane of periodicity) polarizations, respectively. Here we introduce the termination parameter $\tau$ ($0 \leq \tau < 1$), which corresponds to the length of the unit cell in the *z* direction and describes the surface termination. When $\tau = 0$, the outermost row of air holes exhibit a complete profile (i.e., perfect termination, as shown in the inset (1) of Fig. 1) and linearly increase with increasing the height of the surface. As reported in Ref. †, the magnitude of the inverse GH shift of TM and TE polarizations is affected by surface termination. With perfect termination, TM-polarized light is characterized by inverse GH shift of about —24.63*a* and TE-polarized light by a very small inverse GH shift. However, when surface termination of the outermost row of air holes is cut in quarter or so, i.e., τ = 0.25, the inverse GH shift of the TE state presents the maximum value of − 28.27a and the inverse GH shift of the TM state is zero. Furthermore, the effects of surface termination on the inverse GH shift differ in TM and TE polarizations. In order to appreciate the various aspects of inverse GH shift at the interface between air and NRPhC, and to explore the physical mechanism for the difference between TE and TM polarizations, it is worthwhile studying the energy flux around the interface between air and NRPhC. Here, we calculate the energy flux of NRPhC with surface termination τ = 0 and 0.25, because the inverse GH shift of TM and TE polarizations at this two surface terminations are much different, which will be instrumental for getting physical picture of inverse GH shift.

First of all, we calculate the inverse GH shift as a function of the incidence angle with surface termination of τ=0 for TM and TE polarizations. The incident beam used to illuminate the interface between air and NRPhC has Gaussian profile in its cross—section with width of 16 a. In this paper we use FDTD with the boundary condition of PML[9] to calculate field distribution, and use follow formula to extract GH shift[10]:

$$S = \frac{\int_x x \, |E^r(x,0)|^2 \, dx}{\int_x |E^r(x,0)|^2 \, dx} \quad (1)$$

where $E(x,z)$ is the reflected field distribution.

The results are presented in Fig. 1. Expectedly, the maximum value of inverse GH shift for TM polarization is −24.63a, which is much larger than that of −0.5888a of TE polarization. And the profile curves of TM and TE polarizations is Direc-type with HFHW about tens of degrees around the critical angle, which is different to the condition of metal. In the situation of reflection of TM polarized beam on the surface of metal (see the inset (2) of Fig. 1) the value of inverse GH shift exponentially increases with the incidence angle and theoretically gets its maximum value at 90° [11]. In TE polarization normal GH shift with infinitesimal value occurs at the surface of metal.

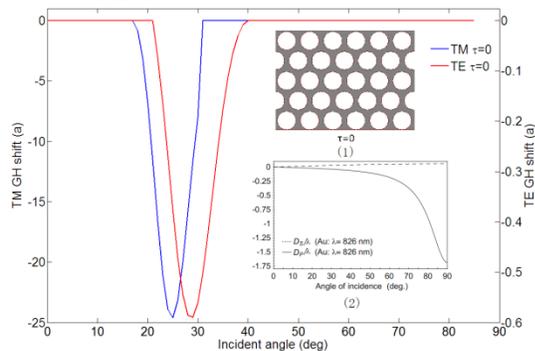

Fig.1.Curves of inverse GH shift of NRPhC with surface termination τ = 0 versus incidence angle for TM and TE polarizations. The parameters of the NRPhC are as follows: $\varepsilon_{slab}$= 12.96, $\varepsilon_{air}$= 1, the radius of air holes is $r$ = 0.4 a, where a is the lattice constant. The incident beam, with frequencies of 0.51 ωa/2πc and 0.33 ωa/2πc for TM and TE polarizations, respectively, has Gaussian profile with width of 16 a.

To better investigate the inverse GH shift of NRPhC, figure 2 gives the energy flux of a TM polarized incident beam illuminating the interface(denoted by white line) between air and NRPhC from lower left at angles of 15°, 20°, 26°, 30°, and 35°, where O is the incident origin (the same below). The color bar maps the intensity of the energy. As expected, negative refraction at the NRPhC–air interface occurs, as indicated by red arrows in Fig. 2(a). Unlike in homogeneous medium, however, the field inside the NRPhC seems to be confined within slant strips oriented along –36.8° to the normal (denoted by short perpendicular magenta line). For better visualization, Fig. 2(b) plots the energy flux around the interface, with about fivefold magnification along both directions (the same for Figs. 2(c)–2(f)). Fig. 2(b) shows that, it is from the dielectric material, represented by dark red regions between air holes (with coordinates in x axis ±n, n the integer) at the interface, that the incident field penetrates into NRPhC. Inside the NRPhC, main of the energy propagate along slant strips (yellow region) consisting of air holes. The dark blue strip-type areas in Fig. 2(b) consist of dielectric gaps, with few field circumscribed there.

Figure 2(c) depicts the energy flux around the interface with the incidence angle of 20°. Similar to the situation of 15°, the field penetrates into NRPhC through the dielectric gaps at the interface, and has main energy confined within the air holes. However, with further observation on the unit cell of NRPhC, we can see that the energy of each air hole not only flows to the upper left hole which is the same as the situation in Fig. 2(b), but flows to the left one as well, indicating the growth of backward wave.

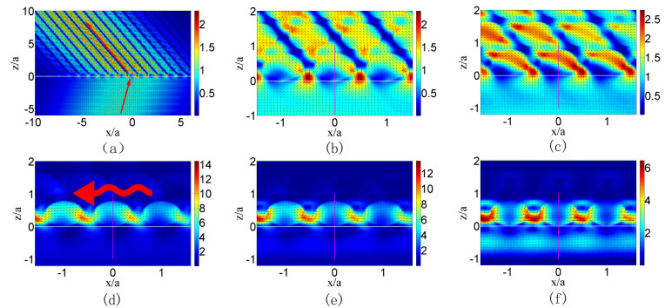

Fig.2. The energy—flux pattern around the NRPhC-air interface (indicated by white line) with incidence angles of (a) 15°, (b) 15°, (c) 20°, (d) 26°, (e) 30°, and (f) 35° for TM polarization. The parameters of NRPhC are same as Fig. 1. The big red arrows indicate the direction of energy flux, and only (a) and (d) are marked with red arrow for neatness, the same as follow.

Figures 2(d)–2(f) give the energy flux around the interface with the incidence angle of 26°, 30° and 35°, respectively. Combined with the situations of Figs.

2(b) and 2(c), we can get a general behavior: with increase of incidence angle, less and less energy of each air hole flow to the upper left one, more and more energy flow to the left one, corresponding to the growth of inverse GH shift. When the incidence angle is exactly equal to the critical angle of 26°, all of the energy inside NRPhC flow toward left, resulting in the maximum value of inverse GH shift. Once the incidence angle outnumbers the critical angle, on one hand, less and less the incident field penetrates into NRPhC (see the cyan region below white line in Fig. 2(f)). On the other hand, the field inside NRPhC gradually forms vortices, shown in Fig. 2(f) is the situation of the incidence angle of 35°.

Compared with Fig. 6 of Ref. 8, we can see that the energy flux of inverse GH shift of NRPhC is different from that of negative–permittivity material (i.e., metal). In negative-permittivity material there are closed-loop fluxes around the interface, irrespective to incidence angle. The reason is that negative–permittivity material supports surface Plasmon wave in TM polarization [12], thus there are closed–loop flux around the interface, irrespective of incidence angle. That is, inverse GH shift of negative–permittivity material originates from surface Plasmon [13]. However, NRPhC supports backward surface wave which is related to incidence angle. From Fig.2 we can find that the inverse GH shift of NRPhC originates from backward surface wave. Besides, for large incidence angle, there are vortices located at $\pm n/2$, $n$ the integer.

For the case of TE state, we also plot the energy flux around the NRPhC–air interface. Figure 3 gives the situation of the incidence angles of 15°, 20°, 29°, 35°, and 40°. We can see that the energy–flux pattern of TE state inside the NRPhC is much different from that of TM state. For better visualization, we amplify the region around the origin with fivefold magnification in Fig. 3(b) (the same for Figs. 3(c)–3(f)). The field inside NRPhC propagates negatively in a zig-zag way. Unlike TM state, main of the energy inside the NRPhC is circumscribed within the dielectric regions, as represented by the curved dark orange strips.

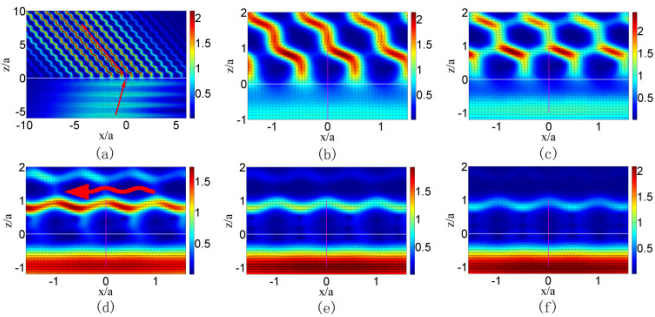

Fig.3. The energy—flux pattern around the NRPhC-air interface (indicated by white line) with incidence angles of (a) 15°, (b) 15°, (c) 20°, (d) 29°, (e) 35°, and (f) 40° for TE polarization. The parameters of NRPhC are same as Fig. 1.

Shown in Figs. 3(c)–3(f) are the situations of the incidence angles of 20°, 29°, 35°, and 40°, respectively. Similar to the situation of TM polarization, with increase of incidence angle, more and more energy transfer into the backward wave of NRPhC, and gets the maximum value at critical angle of 29° and then decreases with the continual increase of incidence angle. However, there are two points should be stressed. First, there is little energy transferred into NRPhC, which is indicated by the color of lower air region in Figs. 3(d)–3(f). Second, the energy of backward wave inside NRPhC is, not concentrated within the outermost row of air holes like the situation of TM state, but concentrated in the region about a lattice beneath the NRPhC surface, as shown in Figs. 3(d)–3(f). The above two points directly results in that the inverse GH shift of TE polarization is very small.

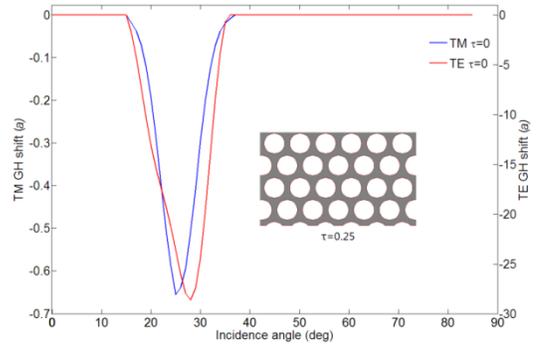

Fig.4. Curves of inverse GH shift of NRPhC with surface termination τ = 0.25 versus incidence angle for TM and TE polarizations. The schematic parameters of NRPhC are same as used in Fig. 1.

As a comparison with the situation of τ=0, we now determine the inverse GH shifts as a function of incidence angle for TM and TE polarizations when surface termination τ=0.25. An inverse result is obtained: large inverse GH shift with the maximum value of –28.27 $a$ is observed for TE polarization, whereas that of TM polarization is zero, as shown in Fig. 4.

Equally, we plot in Figs. 5 and 6 the corresponding energy flux around the interface between NRPhC and air for TM and TE polarizations, respectively. On one hand, in the case of TM polarization, the energy—flux patterns in Fig. 5 are much different to that in Fig. 2. First, main of incident field experiences specular reflection at the interface between air and NRPhC, only little field penetrates into NRPhC, which is clearly indicated by the color of lower air region. Second, the penetrated field inside NRPhC forms vortices at the inner surface of NRPhC, irrespective to incidence angle. That is, no backward surface wave is present at the inner surface of NRPhC when surface termination τ=0.25. Those two facts result in zero inverse GH shift in TM polarization when surface termination τ=0.25. On the other hand, in the case of TE polarization, the energy-flux patterns inside NRPhC with surface termination τ = 0.25 are similar with that of surface termination τ = 0, except for the change of the color of lower air region from red to blue, indicating the growth of the penetrated energy from air to NRPhC. In addition, the backward wave, originally located about a lattice beneath the surface in the situation of perfect termination, is induced to the inner surface of NRPhC (see Figs. 6(d)-6(f)). Therefore, large inverse GH shift about —28.27 $a$ is observed in TE polarization when surface termination τ = 0.25.

On the basis of energy-flux patterns of inverse GH shift of surface termination 0.25, we confirm that the inverse GH shift of NRPhC has different physical mechanism with that of metal. The reasons are as follows: First, in TM polarization there is no closed-loop flux around the interface, but backward surface wave inside inner surface of NRPhC. Second, in TE polarization there is large inverse GH shift resulted by backward surface wave, unlike the infinitesimal small positive GH shift of metal.

Third, the energy-flux pattern of TM polarization is closely related to surface termination, and infinitesimal inverse GH shift is accompanied by vortices in energy-flux pattern in the inner surface of NRPhC. Forth, the energy-flux pattern of TE polarization is not affected by surface termination, but its position in NRPhC is closely related to surface termination. And the inverse GH shift of NRPhC is different to that of metal. Because there is no closed-loop flux around the interface between air and NRPhC in TM polarization; besides, there is large inverse GH shift for TE polarized incident beam at the surface of NRPhC, which is impossible at the surface of metal.

### ACKNOWLEDGMENTS

This work was partially supported by National Basic Research Program of China (2011CB707504), National Natural Science Foundation of China (11104184, 61177043, 61308096), The Innovation Fund Project for Graduate Student of Shanghai (JWCXSL1401).

†. Jinbing Hu, Binming Liang, Jiabi Chen, Qiang Jiang, Yan Wang, Songlin Zhuang, Optics Letters, submitted. ID:231831.

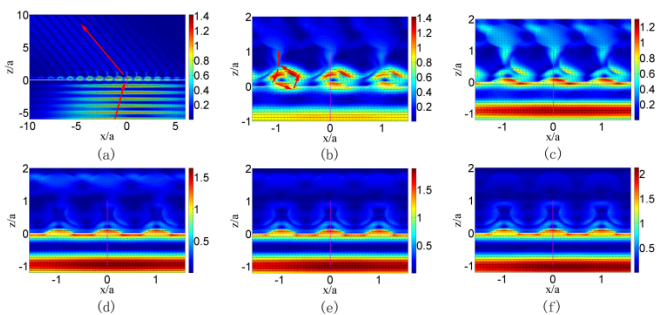

Fig.5. The energy—flux pattern around the NRPhC-air interface (indicated by white line) with incidence angles of (a) 15°, (b) 15°, (c) 20°, (d) 26°, (e) 30°, and (f) 35° for TM polarization. The parameters of NRPhC are same as Fig. 4.

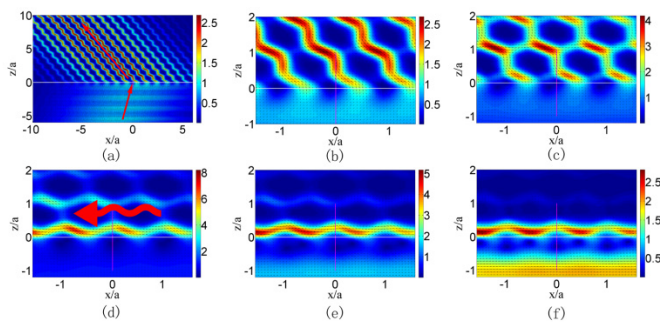

Fig.6. The energy—flux pattern around the NRPhC-air interface (indicated by white line) with incidence angles of (a) 15°, (b) 15°, (c) 20°, (d) 29°, (e) 35°, and (f) 40° for TE polarization. The parameters of NRPhC are same as Fig. 4.

In conclusion, we have investigated the energy—flux pattern of the inverse GH shift of two-dimensional (2D) negatively refractive photonic crystal (NRPhC) consisting of air holes arranged in hexagonal lattice in a dielectric background，some conclusions can be drawn as follows: First, inverse GH shift of NRPhC is resulted by backward surface wave of NRPhC. Second, the magnitude of inverse GH shift is strongly affected by surface termination.